\documentclass[aps,prb,twocolumn,superscriptaddress,showpacs,amsmath]{revtex4}

\usepackage{epsf,latexsym}
\usepackage{amssymb,amsmath}
\usepackage{times}
\usepackage{bm}
\usepackage{verbatim}

\newcommand{\bra}[1]{\langle #1 |}
\newcommand{\ket}[1]{| #1 \rangle}
\newcommand{\braket}[2]{\langle #1 | #2 \rangle}

\newcommand\f{{\mathcal{F}}}

\newcommand\tr{{\mbox{Tr\,}}}

\newcommand{\ignore}[1]{}

\newcommand{\be}{\begin{equation}}
\newcommand{\ee}{\end{equation}}
\newcommand{\ba}{\begin{eqnarray}}
\newcommand{\ea}{\end{eqnarray}}

\def\CC{{\rm\kern.24em \vrule width.04em height1.46ex depth-.07ex
    \kern-.30em C}}
\def\P{{\rm I\kern-.25em P}}

\def\RR{{\rm
         \vrule width.04em height1.58ex depth-.0ex
         \kern-.04em R}}

\def\bbbone{{\mathchoice {\rm 1\mskip-4mu l} {\rm 1\mskip-4mu l}
{\rm 1\mskip-4.5mu l} {\rm 1\mskip-5mu l}}}

\def\bbbc{{\mathchoice {\setbox0=\hbox{$\displaystyle\rm C$}\hbox{\hbox
to0pt{\kern0.4\wd0\vrule height0.9\ht0\hss}\box0}}
{\setbox0=\hbox{$\textstyle\rm C$}\hbox{\hbox
to0pt{\kern0.4\wd0\vrule height0.9\ht0\hss}\box0}}
{\setbox0=\hbox{$\scriptstyle\rm C$}\hbox{\hbox
to0pt{\kern0.4\wd0\vrule height0.9\ht0\hss}\box0}}
{\setbox0=\hbox{$\scriptscriptstyle\rm C$}\hbox{\hbox
to0pt{\kern0.4\wd0\vrule height0.9\ht0\hss}\box0}}}}

\def\bbbz{{\mathchoice {\hbox{$\sf\textstyle Z\kern-0.4em Z$}}
{\hbox{$\sf\textstyle Z\kern-0.4em Z$}}
{\hbox{$\sf\scriptstyle Z\kern-0.3em Z$}}
{\hbox{$\sf\scriptscriptstyle Z\kern-0.2em Z$}}}}

\newcommand{\putfig}[2]{$$\leavevmode\hbox{\epsfxsize=#2 cm
   \epsffile{#1.eps}}$$}

\begin{document}

\title{Quantum fidelity and quantum phase transitions in matrix product states}
\author{Marco Cozzini}
\affiliation{Dipartimento di Fisica, Politecnico di Torino, Corso Duca degli Abruzzi 24, I-10129 Torino, Italy}
\affiliation{Quantum Information Group, Institute for Scientific Interchange (ISI), Viale Settimio Severo 65, I-10133 Torino, Italy}
\author{Radu Ionicioiu}
\affiliation{Quantum Information Group, Institute for Scientific Interchange (ISI), Viale Settimio Severo 65, I-10133 Torino, Italy}
\author{Paolo Zanardi}
\affiliation{Quantum Information Group, Institute for Scientific Interchange
(ISI), Viale Settimio Severo 65, I-10133 Torino, Italy}
\affiliation{Department of Physics and Astronomy, University of Southern 
California, Los Angeles, CA 90089-0484, USA}

\begin{abstract}
Matrix product states, a key ingredient of numerical algorithms widely employed in the simulation of quantum spin chains, provide an intriguing tool for
quantum phase transition engineering. At critical values of the control
parameters on which their constituent matrices depend, singularities in the
expectation values of certain observables can appear, in spite of the
analyticity of the ground state energy.
For this class of generalized quantum phase transitions we test the validity
of the recently introduced fidelity approach, where the overlap modulus of
ground states corresponding to slightly different parameters is considered.
We discuss several examples, successfully identifying all the present
transitions. We also study the finite size scaling of fidelity derivatives,
pointing out its relevance in extracting critical exponents.

\end{abstract}


\maketitle

\section{Introduction}

The study of many-body quantum systems is at the same time a fascinating and
challenging field of research. This is due to the richness of inherently complex
phenomena arising in the presence of a large number of interacting particles,
among which quantum phase transitions (QPTs) occupy a distinguished position.
These transitions take place at zero temperature and are driven by an external
parameter, as for example the magnetic field in superconductors.

Recently it has been shown that the quantum fidelity -- the overlap modulus --
of two finite size ground states corresponding to neighboring control
parameters is a good indicator of QPTs.
Indeed, the fidelity typically drops abruptly at critical points, as a
consequence of the dramatic state transformation involved in a transition.
This approach has been tested in various contexts where the ground state can
be calculated exactly, either analytically or numerically. These include Dicke
model, one-dimensional $XY$ model in a transverse field, and general quadratic
fermionic Hamiltonians \cite{gs_overlap,fid_long}.

In the present article we further pursue this approach by analyzing
QPTs
described by matrix product states (MPSs) \cite{mps_refs}.
These states are at the basis of efficient numerical methods used in the
analysis of spin chain systems, as the density matrix renormalization group
(DMRG) algorithm. When considered dependent on a control parameter $g$ they can
give rise to \textit{generalized} QPTs, i.e., transitions where some observable
quantity presents a non-analytic behavior in spite of the regularity of the
ground state energy \cite{qpt_mps}.
We show that the quantum fidelity of two neighboring (in terms of $g$) MPSs is
an effective tool not only in detecting the critical point $g_c$ but also in
giving the correct scaling at $g_c$.
The success of the fidelity approach in analyzing MPS-QPTs further proves the
generality of the procedure, which, as stressed in
Refs.~\onlinecite{gs_overlap,fid_long},
does not require any a priori understanding of the structure (order parameter,
correlation functions, topology, etc.) of the considered system. It is also
worth pointing out that this method seems to have some advantage with respect
to other quantum information based techniques. For example, quantum phase
transitions in MPSs do not give rise to the logarithmic divergence of the
entropy of block entanglement observed in other systems \cite{block_ent},
thereby ruling out this quantity as a reliable transition indicator.
On the other hand, other entanglement measures can be used, e.g., single site
entanglement and concurrence, whose derivatives often provide interesting
information about criticality \cite{osterloh02}.
We discuss the results
obtained by these methods in one of the simple examples we analyze below.
As it will be shown, also these latter measures turn to be less effective than
the fidelity approach.

In the following we will first present a derivation of the overlap formula for
general MPSs (Sec.~\ref{sec:general}), discussing some general features of
these systems, and then study in detail the MPS examples introduced in
Ref.~\onlinecite{qpt_mps} (Sec.~\ref{sec:examples}).
As in Refs.~\onlinecite{gs_overlap,fid_long}, the analysis will be carried out
also in terms of fidelity derivatives, which are the most suitable tool to
observe finite size scaling properties. In particular, we will explicitly show
how to extract the critical exponent of the correlation length from these
quantities (Subsec.~\ref{subsec:IIordQPTs}), thus demonstrating the
independence and the completeness of our approach.
Finally, we will report on the entanglement analysis in
Subsec.~\ref{subsec:ent}.

\section{Fidelity for matrix product states}
\label{sec:general}

Suppose we have a closed spin chain with $N$ sites. Let $d$ be the dimension of
the Hilbert space at each site. The matrix product states are defined as
\cite{qpt_mps}
\be
\ket{g} := \ket{\psi(g)} = \frac{1}{\sqrt\mathcal{N}}\sum_{i_1,\dots,i_N=0}^{d-1}
\tr(A_{i_1}\dots{}A_{i_N})\ket{i_1\dots{}i_N} \ ,
\ee
where the $A_j$'s, with $j=0,\dots,d-1$, are $D\times D$ matrices, $D$ is the
dimension of the bonds in the so-called valence bond picture, and ${\cal
N}:=\sum_{i_1,\dots,i_N=0}^{d-1}|\tr(A_{i_1}\dots{}A_{i_N})|^2$ a normalization
factor. Since our goal is to explore quantum phase transitions in such states,
we assume that the matrices $A_j=A_j(g)$ depend on one or more parameters,
generically denoted by $g$. The overlap of two MPSs corresponding to different
$g$'s is given by
\ba
\lefteqn{\braket{g_1}{g_2} = [\mathcal{N}(g_1)\mathcal{N}(g_2)]^{-1/2}} \\
&&\times\sum_{\bm{i}\in\bbbz_d^N}\tr[A_{i_1}^*(g_1)\dots{}A_{i_N}^*(g_1)]
\tr[A_{i_1}(g_2)\dots{}A_{i_N}(g_2)] \ , \nonumber
\ea
where $\bm{i}=(i_1,\dots,i_N)$.
In order to simplify the formul\ae, we will work with the unnormalized overlap
\be
F(g_1, g_2) := \sqrt{\mathcal{N}(g_1)\mathcal{N}(g_2)}\,\braket{g_1}{g_2}
\label{overlap}
\ee
and we obviously have $\mathcal{N}(g_a)=F(g_a, g_a)$, $a=1,2$.
Using the identity $\tr(A)\tr(B)=\tr(A\otimes B)$, the overlap can be
computed exactly for MPSs \cite{zhou} and we obtain
\be\label{eq:Fsum_lam}
F(g_1, g_2)= \tr[E^N(g_1,g_2)]= \sum_{k=1}^{D^2} \lambda_k^N(g_1,g_2) \ ,
\ee
where $\lambda_k$ are the eigenvalues of the $D^2\times
D^2$ matrix
\be
E(g_1,g_2) := \sum_{i=0}^{d-1} A_i^*(g_1) \otimes A_i(g_2) \ .
\ee
The latter is the generalization of the transfer operator
$E_\bbbone(g):=E(g,g)$, which we assume to be diagonalizable with eigenvalues
$\lambda_k(g):=\lambda_k(g,g)$.
Also, since $A\otimes{}B$ and $B\otimes{}A$ are isospectral, one has
$F(g_1,g_2)=F(g_2,g_1)^*$.

The main object of our study will be the fidelity of two neighboring states
\be\label{eq:f(g;d)}
\f(g;\delta):= |\braket{g-\delta}{g+\delta}| =
\frac{|F(g-\delta,g+\delta)|}{\sqrt{\mathcal{N}(g-\delta)\mathcal{N}(g+\delta)}}
\ee
for a small variation $\delta$
in the parameter space spanned by $g$. Defined this way, the fidelity depends
not only on the parameter $g$ driving the QPT, but also on its variation
$\delta$.
Expanding the fidelity in $\delta$ we obtain
$\f(g;\delta)\simeq1+
{\partial_\delta^2\f(g;\delta)|}_{\delta=0}\delta^2/2$
(since $\f(g;\delta)$ reaches its maximum for $\delta=0$ the first derivative
vanishes).
Thus the rate of change of the fidelity close to a critical point is given by
the second derivative of the fidelity \cite{gs_overlap,fid_long}
\be
S(g):={\partial^2_\delta\f(g;\delta)|}_{\delta=0}
\ee
and this is the relevant quantity
we will study to determine the scaling at the critical point $g=g_c$.
Note that $S(g)$ is connected to the ground state variation
$\ket{\psi'(g)}=\partial_g\ket{\psi(g)}$.
For example, for real states one has the simple second order expansion
$\f(g;\delta)\simeq1-2\delta^2||\psi'(g)||^2$.
The general expression of the second derivative has the form
\footnote{Considering the fidelity $|\braket{g-\delta}{g+\delta}|$ corresponds
to taking a variation $2\delta$ between the states. This gives rise to the
factor $4$ in formula~(\ref{eq:S(g)general}) and differs from
Refs.~\onlinecite{gs_overlap,fid_long}.}
\be \label{eq:S(g)general}
S(g)= -4\,{\partial_{g_1}\partial_{g_2}\ln{F(g_1,g_2)}|}_{g_1=g_2=g} \ .
\ee
Indeed, from $\f(g;0)=1$ and ${\partial_\delta\f(g;\delta)|}_{\delta=0}=0$
one has
${\partial_\delta^2\ln\f(g;\delta)|}_{\delta=0}={\partial_\delta^2\f(g;\delta)|}_{\delta=0}=S(g)$ and
then
\ba
S(g) & = & \frac{1}{2}\partial_\delta^2
\ln\left.\frac{F(g-\delta,g+\delta)F(g+\delta,g-\delta)}
{F(g-\delta,g-\delta)F(g+\delta,g+\delta)}\right|_{\delta=0} = \nonumber\\
& = & {(\partial_\delta^2-\partial_g^2)\ln{F(g-\delta,g+\delta)}|}_{\delta=0}
\ ,
\ea
from which Eq.~(\ref{eq:S(g)general}) follows immediately.

The general behavior of the fidelity for matrix product states in the
thermodynamic limit (TDL) can be inferred from Eq.~(\ref{eq:Fsum_lam}).
Without loss of generality we can assume the eigenvalues $\lambda_k$ of the
generalized transfer operator $E$ to be ordered $|\lambda_1|\geq|\lambda_2|\geq\ldots\geq|\lambda_{D^2}|$.
Then $F(g_1,g_2)=\sum_{k=1}^{D^2}\lambda_k^N(g_1,g_2)=
\lambda_1^N[1+\sum_{k=2}^{D^2}(\lambda_k/\lambda_1)^N]$ and, provided
$|\lambda_1|>|\lambda_k|$ for $k=2,\dots,D^2$, in the thermodynamic limit
$N\to\infty$ one finds $F(g_1,g_2)\sim\lambda_1^N$.
Since the same holds for the normalization factors $\mathcal{N}(g)=F(g,g)$,
whenever the maximum eigenvalues $\lambda_1$ are non-degenerate, the large $N$
behavior of the fidelity must be of the form
$\mathcal{F}(g;\delta)\sim\alpha^N$, with $0\leq\alpha\leq1$ by construction.
Typically, $\alpha\neq1$ and the fidelity decays exponentially in the
TDL \cite{zhou}.
If $E_\bbbone(g)$ exhibits a level crossing in the largest eigenvalue for a
critical coupling $g_c$, the degeneracy $|\lambda_1(g_c)|=|\lambda_2(g_c)|$
gives rise to a discontinuity of some expectation values in the TDL
(called a \textit{generalized} QPT in Ref.~\onlinecite{qpt_mps}).
In this case the previous discussion has to be modified in order to include all the
degenerate eigenvalues.
In general, the vanishing of the fidelity is typically
strongly enhanced at critical points.

A similar analysis can be done for the second derivative $S(g)$ of the
fidelity. When a single eigenvalue $\lambda_1$ dominates, in the TDL one can
replace $F$ with $\lambda_1^N$ in Eq.~(\ref{eq:S(g)general}), thereby
recovering the expected \cite{fid_long} linear scaling $\propto{N}$:
\be \label{eq:S(l1)}
S(g) \sim -4N\,
{\partial_{g_1}\partial_{g_2}\ln{\lambda_1 (g_1,g_2)}|}_{g_1=g_2=g} \ .
\ee
In contrast, at the critical point two or more eigenvalues of $E(g_1,g_2)$ are
equal in modulus.
For example, at $g=g_c$ one can have
$|\lambda_1(g_c)|=|\lambda_2(g_c)|>|\lambda_k(g_c)|$ for $k>2$. Then in the TDL
\ba \label{eq:S(|l1|=|l2|)}
S(g_c) & \sim & \lim_{g\to{}g_c}-4\,{\partial_{g_1}\partial_{g_2}
\ln(\lambda_1^N+\lambda_2^N)|}_{g_1=g_2=g} = \nonumber\\
& = & \lim_{g\to{}g_c}\left\{
-N^2\frac{4\lambda_1^N\lambda_2^N}{(\lambda_1^N+\lambda_2^N)^2}
\partial_{g_1}\!\ln\frac{\lambda_1}{\lambda_2}
\partial_{g_2}\!\ln\frac{\lambda_1}{\lambda_2}\right. \nonumber \\
&& \!\!\left.\left.-4N\sum_{k=1}^2\frac{\lambda_k^N}{\lambda_1^N+\lambda_2^N}
\partial_{g_1}\partial_{g_2}\!\ln\lambda_k\right\}\!\right|_{g_1=g_2=g} \!\!\ ,
\ea
where $\lambda_{1,2}=\lambda_{1,2}(g_1,g_2)$.
Hence, at the transition one typically recovers the divergence
$S(g_c)\propto{}N^2$ already observed in Ref.~\onlinecite{fid_long}.
We also note that, by defining $\theta:=\ln(\lambda_1/\lambda_2)$, the term in
the second line of Eq.~(\ref{eq:S(|l1|=|l2|)}) can be rewritten in the compact
form $[N|\partial_{g_1}\theta|/\cosh(N\theta/2)]^2$.
This follows from the equality $\partial_{g_1}\ln(\lambda_1/\lambda_2)
\partial_{g_2}\ln(\lambda_1/\lambda_2)=
|\partial_{g_1}\ln(\lambda_1/\lambda_2)|^2$, due to the relation
$\lambda_{1,2}(g_2,g_1)=\lambda_{1,2}^*(g_1,g_2)$.
Then, it is clear that this term vanishes for $N\to\infty$ unless $\Re\theta=0$,
i.e., $|\lambda_1|=|\lambda_2|$.
In the latter case, instead, assuming $\partial_{g_1}\theta$ to be finite, the
scaling $\propto{}N^2$ is evident.
Finally, we point out that the limit $g\to{}g_c$ in Eq.~(\ref{eq:S(|l1|=|l2|)})
is necessary to deal with possible singularities in the derivatives of the
$\lambda_k$'s at the transition (see the examples below).

\section{Examples}
\label{sec:examples}

In the following we are going to study the explicit examples of MPSs described
in Ref.~\onlinecite{qpt_mps}. Without loss of generality we assume
$g_1=g-\delta<g_2=g+\delta$.

\subsection{Fidelity and second order MPS-QPTs}
\label{subsec:IIordQPTs}

\begin{figure}
\putfig{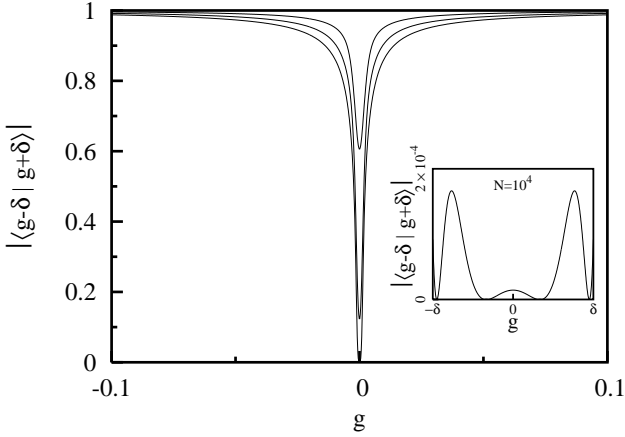}{8.6}
\caption{Fidelity $\f(g;\delta)=|\braket{g-\delta}{g+\delta}|$ as a function of
$g$ for Example 1 with $\delta=10^{-3}$ and $N=1000,2000,3000$.
Inset: close to the critical point, the fidelity oscillates for states
belonging to different phases ($\delta=10^{-3}$, $N=10^4$).}
\label{fig:fid2body}
\end{figure}

{\em Example 1}. Take $D=2, d=3$ and $\{ A_j \}=\{-Z,\sigma_-,g\sigma_+\}$; in
the following we will use the notation
$X,Z$ for the Pauli matrices, $\sigma_\pm$ being the corresponding raising and
lowering operators.
This one-parameter family of MPSs contains the ground state of spin-1 AKLT model
for $g=\pm2$ and has a critical point at $g_c=0$ (see Ref.~\onlinecite{qpt_mps}).
The model is also significant because it has non-local string order. The parent
Hamiltonian with two-body interaction is given by
\ba
H & = & \textstyle\sum_i[(2+g^2)\bm{S}_i\cdot\bm{S}_{i+1}+
2(\bm{S}_i\cdot\bm{S}_{i+1})^2+ \nonumber\\
&&+2(4-g^2)(S_i^z)^2+(g+2)^2(S_i^zS_{i+1}^z)^2+ \nonumber\\
&&+g(g+2)\{S_i^zS_{i+1}^z,\bm{S}_i\cdot\bm{S}_{i+1}\}] \ ,
\ea
where $\bm{S}_i$ is the spin operator acting on the $i$-th site.
The eigenvalues of the transfer operator $E= Z\otimes Z+ \sigma_- \otimes
\sigma_- + g_1g_2 \sigma_+ \otimes \sigma_+$ are $( -1, -1, 1\pm \sqrt{g_1 g_2}
)$ and Eq.~(\ref{eq:Fsum_lam}) yields
\be
F(g_1,g_2)= (1+\sqrt{g_1 g_2})^N+(1-\sqrt{g_1 g_2})^N+2(-1)^N
\ee
and ${\cal N}(g)=F(g,g)= (1+g)^N+ (1-g)^N+ 2(-1)^N$.

Let us analyze the behavior of the fidelity $\f(g;\delta)$ defined in
Eq.~(\ref{eq:f(g;d)}). Since $\f(g;\delta)$ is evidently symmetric in
both $g$ and $\delta$, we can safely assume $g\geq0$, $\delta>0$.
Three cases can then be distinguished, according to the
behavior of the eigenvalues of the transfer operators:

(i) $g>\delta$, where $\f(g;\delta)\sim\alpha^N$ for $N\gg1$, with
$\alpha=(1+\sqrt{g^2-\delta^2})^2/\sqrt{(1+g-\delta)(1+g+\delta)}<1$;

(ii) $g=\delta$, where $\f(\delta;\delta)\sim{}a/(1+2\delta)^{N/2}$ for
$N\gg1$, with $a=2$ for $N$ even and $a=2\delta\sqrt{N(N-1)}$ for $N$ odd (this
is the case where the largest eigenvalue of $E(g-\delta,g+\delta)$ is
degenerate and the decaying behavior of the fidelity is slightly modified);

(iii) $g<\delta$, where two of the eigenvalues of $E(g_1,g_2)$ are complex and,
for $N\gg1$, the fidelity exhibits an oscillatory behavior driven by
$|\cos(N\varphi)|$, $\varphi=\arctan\sqrt{\delta^2-g^2}$, with a decaying
envelope given by $2\alpha^N$,
$\alpha=\sqrt{(1+\delta^2-g^2)/[(1+\delta-g)(1+\delta+g)]}<1$.
Explicitly, at the critical point $g=g_c=0$ the overlap becomes
\be
\braket{-\delta}{\delta} = \frac{(1+\delta^2)^{N/2} \cos(N\varphi_0)+ (-1)^N}
{[(1+\delta)^N+ (1-\delta)^N]/2+ (-1)^N} \ ,
\ee
where $\varphi_0=\arctan\delta$. Then
$\f(0;\delta)=|\braket{-\delta}{\delta}|$ (note that the overlap can be
negative at this point).

\begin{figure}
\putfig{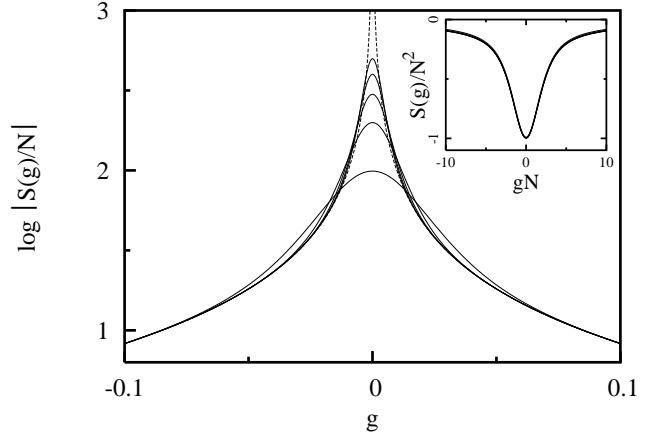}{8.6}
\caption{Finite size scaling for $S(g)$. The second derivative $S(g)$ is
plotted for $N=100,200,\dots,500$ (logarithm is base 10). The peak at $g=0$ scales with $N$
accordingly to Eq.~(\ref{eq:S(0)2body}). The dashed line corresponds to the
asymptotic behavior given by Eq.~(\ref{eq:Slim2body}).
The inset shows the data collapsing for the same curves ($N=100,200,\dots,500$)
when plotted in rescaled units.}
\label{fig:fssS2body}
\end{figure}

The fidelity behavior in the neighborhood of the critical point is shown
in Fig.~\ref{fig:fid2body} for $\delta=10^{-3}$ and various values of $N$.
It is evident that, although for a fixed $\delta$ the fidelity vanishes in the
TDL for any $g$, the rate of decrease is much faster at the transition.
This fidelity drop provides a useful finite size precursor of the quantum
phase transition taking place in the infinite system.
In the figure inset we zoom on the oscillatory behavior in the interval
$g\in(-\delta,\delta)$, where states on different sides of the transition are
considered. Note that oscillations start to be visible only for $N\varphi>\pi$,
i.e., $N>N_{\mathrm{osc}}$, $N_{\mathrm{osc}}=\pi/\arctan\sqrt{\delta^2-g^2}$,
which for $g=0$ and $\delta\ll1$ reduces to $N_{\mathrm{osc}}\sim\pi/\delta$.

The second derivative $S(g)$ of the fidelity can be calculated analytically
using formula (\ref{eq:S(g)general}). We find
\ba
\label{eq:Slim2body}
S(g\neq0) & \sim & -\frac{N}{|g|(1+|g|)^2} \ , \\
\label{eq:S(0)2body}
S(g=0) & = &
\left\{\begin{array}{lc}
-N(N-1) & \mbox{($N$ even)} \\
-(N-2)(N-3)/3  & \mbox{($N$ odd)}
\end{array}\right. \ ,
\ea
where Eq.~(\ref{eq:Slim2body}) corresponds to Eq.~(\ref{eq:S(l1)}) and gives the
asymptotic behavior for large $N$, while Eq.~(\ref{eq:S(0)2body}) is exact.
As in Ref.~\onlinecite{fid_long} the critical point is identified by a
divergence $\propto{}N^2$ of the second derivative $S(g)$ of the fidelity.

In Fig.~\ref{fig:fssS2body} we plot $\log|S(g)|$ for different values of $N$. In the inset of the
figure the curves are plotted in rescaled units, giving rise to data collapsing
(lines for different values of $N$ are practically indistinguishable in the
inset, merging into a single thick line). This shows that $S(g)/N^2$ is
basically a function of $gN$ only.
This is reminiscent of the usual scaling behavior of diverging observables in
second order quantum phase transitions.
Indeed, if an observable $P$ of a one-dimensional system diverges algebraically
in the TDL for some critical coupling $g_c$, i.e.,
$P_\infty\sim{}c_\infty|g-g_c|^{-\rho}$, the finite size scaling Ansatz in the
critical region reads $P_N=N^{\rho/\nu}Q(N|g-g_c|^\nu)$, where $\nu$ is the
critical exponent of the bulk correlation length $\xi_c$ and $Q$ is some
function \cite{fss}.
This is due to the fact that at criticality, once properly rescaled by some
power of the size of the system, the observable is expected to depend only on
the dimensionless ratio $L/\xi_c$ between the system size $L\propto{}N$ and the
correlation length.
Since the latter diverges at the transition like $\xi_c\propto1/|g-g_c|^\nu$
one immediately finds $L/\xi_c\propto{}N|g-g_c|^\nu$.
In our case $P_N=S/N=NQ(N|g|)$ with $g_c=0$, so that we obtain $\rho=\nu=1$.
The result $\rho=1$ agrees with Eq.~(\ref{eq:Slim2body}), which,
for $g\to0$, yields $P_\infty\sim{}-|g|^{-1}$.
The value of $\nu$ can instead be checked by explicitly calculating the
correlation length
\footnote{The correlation length in MPSs is usually given by the formula
$\xi_c=1/\ln|\lambda_1/\lambda_2|$, where $\lambda_1$
($\lambda_2$) is the eigenvalue of $E_\bbbone(g)$ of (second) largest modulus.
In this case, however, the degeneracy of $\lambda_2=\lambda_3=-1$ does not allow
to apply this expression.},
where one finds $\nu=1$, confirming the finite size scaling estimate.

The next examples of this subsection correspond to three-body interaction
Hamiltonians, again taken from Ref.~\onlinecite{qpt_mps}.
Qualitatively, they all feature the same behavior as Example 1:
in the asymptotic limit $|S(g)|/N$ diverges proportionally to $1/|g|$ at the
critical point $g_c=0$, while the peak height of $S(g=0)$ scales with $N^2$.

{\em Example 2}. Consider $D=d=2$ and
$A_0=(\bbbone-Z)/2+\sigma_-$, $A_0=(\bbbone+Z)/2+g\sigma_+$.
The three-body Hamiltonian is $\bbbz_2$ symmetric and has a critical point at
$g_c=0$, where the state is a Greenberger-Horne-Zeilinger (GHZ) state, and reads
\be
H = \sum_i 2(g^2-1) Z_i Z_{i+1}- (1+g)^2 X_i +
(g-1)^2 Z_i X_{i+1} Z_{i+2} \ .
\ee
The generalized transfer operator $E(g_1,g_2)$ has eigenvalues
$(0,0,1\pm\sqrt{g_1 g_2})$, very similar to Example 1. The function
$F(g-\delta,g+\delta)$ is then still symmetric in $g$ and $\delta$.

Since the largest eigenvalue $\lambda_1$ is the same as in Example 1, the TDL
behavior of $\f(g,\delta)$ and $S(g)$ is unchanged.
However, the parity dependent term $(-1)^N$ is now absent.
Then, the asymptotic behavior of $S(g\neq0)$ is still given by
Eq.~(\ref{eq:Slim2body}), while
the parity dependence of Eq.~(\ref{eq:S(0)2body}) is lost and one simply has
$S(g=0)=-2N(N-1)$.
The latter results follows from Eq.~(\ref{eq:S(|l1|=|l2|)}), which applies
exactly to this case even for finite $N$. Indeed here only two eigenvalues of
$E(g_1,g_2)$ are different from zero and one has the level crossing
$\lambda_1(g=0)=\lambda_2(g=0)$.
Note that, due to the divergence of the double derivative
$\partial_{g_1}\partial_{g_2}\ln\lambda_{1,2}$ calculated in $g_1=g_2=0$, also the
term in the last line of Eq.~(\ref{eq:S(|l1|=|l2|)}) gives rise to a
contribution $\propto{}N^2$, in spite of its apparent linearity in $N$ for
$|\lambda_1|=|\lambda_2|$. This explains the necessity of the limit $g\to g_c$
in Eq.~(\ref{eq:S(|l1|=|l2|)}).

{\em Example 3}. Consider $A_0=X$, $A_1=\sqrt{g}(\bbbone- Z)/2$. The
eigenvalues of $E(g_1,g_2)$ are
$(\pm1,(\sqrt{g_1g_2}\pm\sqrt{g_1g_2+4})/2)$.
As before, the scaling of the system can be seen from the behavior of $S(g)$
in the thermodynamic limit,
driven by $\lambda_1(g_1,g_2)=(\sqrt{g_1g_2}+\sqrt{g_1g_2+4})/2$.
We find
\ba
\label{eq:Slim3body}
S(g\neq0) & \sim & -\frac{4N}{|g| (g^2+4)^{3/2}} \ , \\
\label{eq:S(0)3body}
S(g=0) & = &
\left\{\begin{array}{lc}
-N^2/4 & \mbox{($N$ even)} \\
-(N^2-1)/6  & \mbox{($N$ odd)}
\end{array}\right. \ ,
\ea
where Eq.~(\ref{eq:Slim3body}) gives the large $N$ behavior.
The parity dependence of Eq.~(\ref{eq:S(0)3body}) is caused by the negative
eigenvalues of $E(g_1,g_2)$, giving rise to terms oscillating like $(-1)^N$.

{\em Example 4}. Take $A_0= \sigma_+$, $A_1=\sigma_- + \sqrt{g}(\bbbone+ Z)/2$.
This MPS is the ground state of the following Hamiltonian:
\ba
\nonumber
H&=& -\sum_i g(X_i+ X_i Z_{i+1}+ Z_i X_{i+1}+ Z_i X_{i+1} Z_{i+2}) \\
&+& (1+2 g^2) Z_i- 2Z_i Z_{i+1}- Z_i Z_{i+1} Z_{i+2} \ .
\ea
The eigenvalues of $E(g_1,g_2)$ are
$(0,0,(\sqrt{g_1g_2}\pm\sqrt{g_1 g_2+4})/2)$.
Hence, the asymptotic behavior of $S(g)$ is again given by
Eq.~(\ref{eq:Slim3body}), while $S(0)=-N^2/2$ for $N$ even and
$S(0)=-(N^2-1)/6$ for $N$ odd. Parity dependence arises from
$(\sqrt{g_1g_2}-\sqrt{g_1g_2+4})/2=-1$ for $g_1=g_2=0$.

\subsection{Fidelity, concurrence, and single site entanglement}
\label{subsec:ent}

While all the above examples fit in the same picture, which is typical of
the second order quantum phase transitions already studied in
Refs.~\onlinecite{gs_overlap,fid_long},
the next example, albeit trivial as matrix product state, features a different
behavior, which will serve as a basis for some additional comments on the
fidelity approach to QPTs.

{\em Example 5}. Take $D=d=2$,
$A_0=\begin{pmatrix}1 & 0 \cr 0 & 1+g \end{pmatrix}$,
$A_1=\begin{pmatrix}g^n & 0 \cr 0 & 0 \end{pmatrix}$.
The matrix $E(g_1,g_2)$ has the eigenvalues $1+g_1$, $1+g_2$, $(1+g_1)(1+g_2)$,
and $1+g_1^n g_2^n$.
As discussed in Sec.~\ref{sec:general}, QPTs take place when two or more
eigenvalues of $E_\bbbone(g)=E(g,g)$ share the largest modulus.
We then have to compare $1+g$, $(1+g)^2$, and $1+g^{2n}$.
Clearly, for $g<0$ one has $\lambda_1(g)=1+g^{2n}$, while for
$g\geq0$ the dominant eigenvalue is determined by the ratio
$r(g):=(1+g^{2n})/(1+g)^2$. The equation $r(g)=1$ can be rewritten as
$g(g^{2n-1}-g-2)=0$. Apart from the trivial solution $g_c=0$, for $n\geq2$ a
second solution $g_c'>1$ exists, so that in the region $0\leq{}g\leq{}g_c'$ one
has $\lambda_1(g)=(1+g)^2$, while for $g>g_c'$ one obtains again
$\lambda_1(g)=1+g^{2n}$. For $n=1$ the point $g_c'$ can be considered
at infinity and the latter level crossing never occurs.

In order to evaluate the fidelity we have to substitute the above eigenvalues in
Eq.~(\ref{eq:f(g;d)}). 
For simplicity, here we restrict our analysis to the fidelity of ground states
belonging to the same phase, avoiding to discuss the effects shown in the inset
of Fig.~\ref{fig:fid2body} for Example 1.
In the phase corresponding to the interval $g\in(0,g_c')$, where
$\lambda_1(g)=(1+g)^2$, one has $\lambda_1(g_1,g_2)=(1+g-\delta)(1+g+\delta)$.
The normalized fidelity in the TDL then results
$\f(g;\delta)\sim
[(1+g-\delta)(1+g+\delta)/\sqrt{(1+g-\delta)^2(1+g+\delta)^2}]^N=1$,
so that one has a constant phase in this region, due to the factorization of
the largest eigenvalue of $E(g-\delta,g+\delta)$.
Outside this region such a factorization is absent and the fidelity
vanishes exponentially with the size of the system.

The asymptotic behavior of the second derivative $S(g)$ can be calculated
accordingly.
In the interval $g\in(0,g_c')$, where the eigenvalue $(1+g)^2$ dominates, one
finds $S(g)=0$ in the TDL, while for $g\notin[0,g_c']$, where the leading
eigenvalue is given by $1+g^{2n}$, Eq.~(\ref{eq:S(l1)}) yields the formula
\be\label{eq:Sextriv}
S(g)\sim-4Nn^2g^{2n-2}/(1+g^{2n})^2 \ .
\ee
The latter function, which is evidently symmetric in $g$, has a minimum at
$g_{\mathrm{min}}=-[(n-1)/(n+1)]^{1/2n}$ (since $|g_{\mathrm{min}}|<g_c'$, the
minimum position for $g>0$ is instead given by $g_c'$ itself, as $S(g)$
increases monotonically for $g>g_c'$).
Note that, while $S(g)$ is continuous at $g=g_c=0$, where $S(0)=0$, at the second
critical point $g=g_c'$ one has the discontinuity
$\lim_{g\to(g_c')^-}S(g)=0\neq
\lim_{g\to(g_c')^+}S(g)=-4Nn^2(g_c'+2)/[g_c'(1+g_c')^4]$,
where the latter results is obtained by substituting the defining relation
$r(g_c')=1$ into Eq.~(\ref{eq:Sextriv}).
Furthermore, exactly at the critical point $g_c'$ one has the superextensive
scaling $S(g_c')\sim-N^2[2n-1+(n-1)g_c']^2/(1+g_c')^4$.

\begin{figure}
\putfig{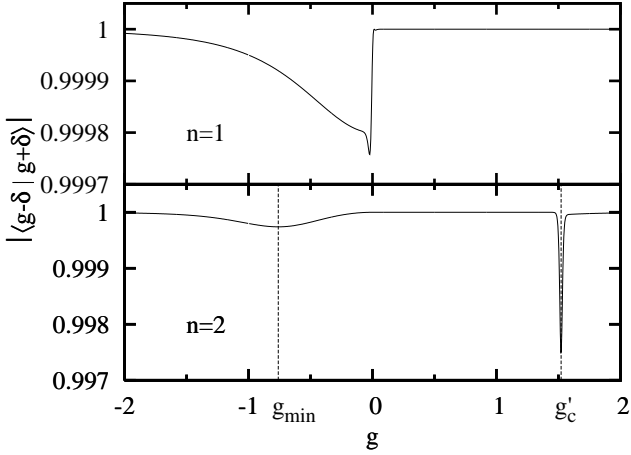}{8.6}
\caption{Fidelity for the model given in Example 5, with $\delta=10^{-3}$ and
$N=100$; $n=1$ (top) and $n=2$ (bottom).}
\label{fig:triv_ex}
\end{figure}

The finite size behavior of the fidelity is shown in Fig.~\ref{fig:triv_ex} for
$N=100$ and $\delta=10^{-3}$. Due to the approximated relation
$\f(g;\delta)\simeq1+S(g)\delta^2/2$ one can there recognize also the scaling of
$S(g)$. In particular, in the lower panel where the case $n=2$ is plotted, the
minimum at $g_{\mathrm{min}}$ and the superextensive scaling at
$g_c'(n=2)\simeq1.521$
are evident.

For this simple example it is also possible to find a compact analytic expression
for the ground state $\ket{g}$. Due to the commutativity $[A_0, A_1]=0$, one
has $\sqrt\mathcal{N}\ket{g}=\sum_{\bm{i}\in\bbbz_2^N}
\mathrm{Tr}(A_0^{N-k}A_1^k)\ket{i_1\dots{}i_N}$ with $k=\sum_{j=1}^Ni_j$. Then,
by using $\mathrm{Tr}(A_0^N)=1+(1+g)^N$ and $\mathrm{Tr}(A_0^{N-k}A_1^k)=g^{nk}$
for $k\neq0$, one finds
\ba
\sqrt\mathcal{N}\ket{g} & \!=\! &
\left[1+(1+g)^N\right]\!\ket{0}^{\otimes{}N}\!+\!
\sum_{k=1}^Ng^{nk}{N \choose k}^{\frac{1}{2}}\ket{D_N^{(k)}} = \nonumber \\
& \!=\! & (1+g)^N\ket{0}^{\otimes{}N}+(1+g^{2n})^{N/2}\ket{\phi(g)}^{\otimes{}N}
\ , \label{eq:gs_ex5}
\ea
where $\ket{D_N^{(k)}}={N \choose k}^{-\frac{1}{2}}
\sum_{\bm{i}\in{}I_k}\ket{i_1\dots{}i_N}$ with
$I_k=\{\bm{i}\in\bbbz_2^N|\sum_{j=1}^Ni_j=k\}$ is a Dicke state \cite{dicke54}
and $\ket{\phi(g)}=(\ket{0}+g^n\ket{1})/\sqrt{1+g^{2n}}$ is a normalized single
site state.
Since the normalization is $\mathcal{N}=\tr{E_\bbbone^N}$,
depending on the dominating eigenvalue of $E_\bbbone$ three cases are
possible in the TDL:

(i) $\lambda_1(g)=(1+g)^2>\lambda_2(g)$ and
$\ket{g}\to\ket{0}^{\otimes{}N}$;

(ii) $\lambda_1(g)=1+g^{2n}>\lambda_2(g)$ and
$\ket{g}\to\ket{\phi(g)}^{\otimes{}N}$;

(iii) $\lambda_1(g)=\lambda_2(g)$ and, in
the non-trivial case $g=g_c'$, one has
\footnote{Note that $\ket{0}^{\otimes{}N}$ and $\ket{\phi(g_c')}^{\otimes{}N}$
become orthogonal in the TDL, in accordance with the normalization factor
$1/\sqrt2$.}
$\ket{g}\to(\ket{0}^{\otimes{}N}+\ket{\phi}^{\otimes{}N})/\sqrt2$.

The above analysis can be summarized in the following table, where we recall the
results for the largest transfer operator eigenvalue $\lambda_1(g)$ and for the
large $N$ behavior of the function $S(g)$ and the ground state $\ket{g}$:
\begin{center}
\begin{tabular}{|c|c|c|c|}
\hline
& $g\in[0,g_c')$ & $g=g_c'$ & $g\notin[0,g_c']$ \\
\hline\hline
$\lambda_1(g)$ & $(1+g)^2$ & $1+g^{2n}=(1+g)^2$ & $1+g^{2n}$ \\
\hline
$S(g)$ & 0 & $\propto{}N^2$ &
$\propto{}N$ \\
\hline
$\ket{g}$  & $\ket{0}^{\otimes{}N}$ &
$(\ket{0}^{\otimes{}N}+\ket{\phi(g_c')}^{\otimes{}N})/\sqrt2$ &
$\ket{\phi(g)}^{\otimes{}N}$ \\
\hline
\end{tabular}
\end{center}

It is worth noting that the apparent drop of the fidelity at $g_{\mathrm{min}}$
in the $g<0$ phase for $n\geq2$ does not correspond to a QPT. Indeed, the
scaling of $S(g_{\mathrm{min}})$ does not present any peculiar behavior with
respect to the rest of the $g<0$ phase.
On the contrary, the scaling of $S$ changes its nature at $g=0$ and $g=g_c'$,
highlighting the transitions, although in a very different way from the typical
behavior observed in the previous second order QPTs.
The transition at $g=0$ corresponds to a discontinuity in
$\partial_g^n\langle{X}\rangle$ (Ref.~\onlinecite{qpt_mps}), while that at
$g=g_c'$ to a discontinuity of $\langle{X}\rangle$ itself
\footnote{\protect In the limit $N\to\infty$ one has
$\langle{X(g)}\rangle\to0$ for $g\in[0,g_c')$,
$\langle{X(g_c')}\rangle\to(g_c')^n/[1+(g_c')^{2n}]$,
and $\langle{X(g)}\rangle\to2g^n/(1+g^{2n})$ for $g{\protect\notin}[0,g_c']$.
Hence, for example, $\lim_{g\to0^-}\partial_g^n\langle{X}\rangle=2n!$, while
$\lim_{g\to0^+}\partial_g^n\langle{X}\rangle=0$.}.
Due to the permutation symmetry arising from the commutativity of $A_0$, $A_1$,
the correlation length cannot here be defined (the usual formula
$\xi_c=1/\ln|\lambda_1/\lambda_2|$ does not hold).
In practice, the transitions of Example 5 appear to be of first order.

For this example we also investigate concurrence and single site
entanglement, mainly focusing on the critical point $g_c=0$.
Of course, the fact that the state is factorized in the TDL trivially implies
the vanishing of these quantities in this limit.
We are however interested in the finite size scaling of the derivatives
of these measures, in the spirit of Ref.~\onlinecite{osterloh02}.
We are motivated by various aspects.
First, as mentioned before, in MPS-QPTs the block entanglement entropy does not
exhibit the logarithmic divergence observed in other quantum phase transitions,
posing the question whether other entanglement properties usually signaling
quantum criticality behave differently in this case.
Second, we are interested in comparing the effectiveness of the fidelity
approach against other quantum information-theoretical methods.
The simple nature of this example allows for a fully analytical treatment.

{\em Concurrence.}
We compute the concurrence for the reduced density matrix of 2 qubits in
the chain (the choice of the qubits is unimportant, due to the permutation
symmetry mentioned above). The density matrix of two spins $\rho^{(2)}$ is
obtained by tracing the initial $\rho=\ket{g}\bra{g}$ over all the other spins
in the chain.
By exploiting the symmetries arising from the commutation relations $[A_i,A_j]=0$
and recalling that
$\rho^{(2)}=\sum_{i,j,k,l=0}^1\rho_{ij,kl}^{(2)}\ket{ij}\bra{kl}$ is a $4\times4$
real symmetric matrix, one finds
$\rho^{(2)}_{ij,kl}=\rho^{(2)}_{ji,kl}=\rho^{(2)}_{ij,lk}$. Together with the
normalization $\tr\rho^{(2)}=1$ this implies that only 5 entries of
$\rho^{(2)}$ are independent, e.g., $\rho_{00,00}^{(2)}$, $\rho_{00,01}^{(2)}$,
$\rho_{00,11}^{(2)}$, $\rho_{01,01}^{(2)}$, $\rho_{01,11}^{(2)}$,
significantly simplifying the calculation in the standard basis.
Otherwise, one can use Eq.~(\ref{eq:gs_ex5}), easily obtaining
\ba \label{eq:rho2}
\rho^{(2)} &=& [(1+g)^{2N}\ket{00}\bra{00}+
(1+g^{2n})^{N}\ket{\phi\phi}\bra{\phi\phi} \\
&& +(1+g)^N(1+g^{2n})(\ket{00}\bra{\phi\phi}+\ket{\phi\phi}\bra{00})]
/\mathcal{N}(g) \ . \nonumber
\ea
Having $\rho^{(2)}$, the concurrence for two qubits is defined as \cite{C}
$C=\max\{0,\sqrt{\mu_1}-\sqrt{\mu_2}-\sqrt{\mu_3}-\sqrt{\mu_4}\}$, where
$\mu_i$'s are the eigenvalues, in decreasing order, of the operator
$\rho^{(2)}(\sigma_y \otimes \sigma_y)(\rho^{(2)})^*(\sigma_y \otimes \sigma_y)$.
We finally have:
\be
C(g)= \frac{2 g^{2n} |(1+g)^N| } {\mathcal{N}(g)} \ .
\ee
In the thermodynamic limit both $C(g)$ and its first derivative vanish, so that the kind of
transition signatures found, e.g., in Ref.~\onlinecite{osterloh02} are absent here.
However, for $n=1$, the fourth derivative $\partial_g^4{C|}_{g=0}$ has a
divergence in the TDL; for $n\geq2$ the singularity is present in even higher
derivatives.

{\em Single site entanglement.}
This quantity is
the von Neumann entropy of a single spin ${\cal
S}=-\mathrm{Tr}(\rho^{(1)}\log_2\rho^{(1)})$, where $\rho^{(1)}$ is the reduced
density matrix of a single site.
Again from Eq.~(\ref{eq:gs_ex5}) one finds
\ba
\rho^{(1)} &=& [(1+g)^{2N}\ket{0}\bra{0}+
(1+g^{2n})^{N}\ket{\phi}\bra{\phi} \\
&& +(1+g)^N\sqrt{1+g^{2n}}(\ket{0}\bra{\phi}+\ket{\phi}\bra{0})]
/\mathcal{N}(g) \ , \nonumber
\ea
whose eigenvalues are $\lambda_\pm=(1\pm\sqrt{1-4\det\rho^{(1)}})/2$, with
$\lambda_+=1-\lambda_-$ and
\be\label{eq:det(rho1)}
\det\rho^{(1)}=\frac{g^{2n}(1+g)^{2N}[(1+g^{2n})^{N-1}-1]}{\mathcal{N}^2(g)} \ .
\ee
The single site entanglement entropy
$\mathcal{S}=-\lambda_+\log_2\lambda_+-\lambda_-\log_2\lambda_-$ vanishes in the
thermodynamic limit at $g=0$, together with its first derivative.
For $n=1$, one finds a divergence in $\partial_g^4{\mathcal{S}|}_{g=0}$, while the
divergence is shifted to higher derivatives for $n\geq2$, similarly to the
concurrence. The considered divergence is however due to the functional form of
the von Neumann entropy. Indeed, $\lambda_-\to0$ for $g\to0$, giving rise to a
singularity in the logarithm.

We conclude this analysis by briefly discussing the TDL of these entanglement
measures at the critical point $g=g_c'$, where the state is not factorized
(see the table above).
The single site reduced density matrix in the TDL is simply
$\rho^{(1)}(g_c')=(\ket{0}\bra{0}+\ket{\phi}\bra{\phi})/2$ and
$\det\rho^{(1)}(g_c')=[1-1/(1+g_c')^2]/4$.
Then $\lambda_\pm(g_c')=[1\pm1/(1+g_c')]/2$ and $\mathcal{S}(g_c')\neq0$.
Note however that $C(g_c')=0$ in the TDL, reminiscent of the behavior of the
GHZ state.
This can be seen from the simple TDL of Eq.~(\ref{eq:rho2}), namely,
$\rho^{(2)}(g_c')=(\ket{00}\bra{00}+\ket{\phi\phi}\bra{\phi\phi})/2$.

\section{Conclusions}

In conclusion we have investigated the quantum fidelity of slightly different
matrix product states dependent on a parameter $g$ in the context of quantum
phase transitions. 
For generic MPSs, the overlap can be related analytically to the eigenvalues of a
generalized transfer operator, among which the largest in modulus plays a crucial
role. If the latter is non-degenerate (in modulus), the fidelity typically
exhibits an exponential decay in the thermodynamic limit and its second
derivative is proportional to the size $N$ of the system.
If instead more eigenvalues share the largest modulus, a quantum phase transition
can take place and the fidelity second derivative $S(g)$ generally scales with
$N^2$.
We have demonstrated this behavior in the exhaustive analysis of some simple
examples, taken from Ref.~\onlinecite{qpt_mps}, where possible exceptions have
also been pointed out.
Moreover, we have shown that the second derivative of the fidelity is a useful
quantity for the quantitative analysis of the scaling properties of the system at
the transition point. From the finite size scaling behavior of $S(g)$ we have
indeed estimated the critical exponent for the correlation length, finding
agreement with the explicit calculation.
Finally, for one of the considered examples, we have analyzed both concurrence and 
single site entanglement.
Although the latter quantities did provide signatures of the undergoing quantum
phase transitions, this information was hidden in high order derivatives, making
it much more difficult to extract than the fidelity (or its second derivative).

The fidelity analysis has the advantage of providing a unified framework to
detect very different types of phase transitions.
For the concrete examples analyzed here, singularities in the fidelity are
related to level crossings in the transfer operator, recovering the known
transition mechanism for matrix product states.
The results presented here further contribute to demonstrate the generality of
the fidelity approach to quantum phase transitions \cite{FiniteT}, supporting
the findings of Refs.~\onlinecite{gs_overlap,fid_long}.

\acknowledgments

We thank L.~Amico and P.~Giorda for valuable discussions and comments.


\end{document}